\documentclass[aps,prb,twocolumn,superscriptaddress,showpacs]{revtex4-1}

\usepackage[pdftex]{graphicx} 
\usepackage{amsmath} 
\usepackage{amssymb} 
\usepackage{mathtools}
\usepackage{color} 
\usepackage{textcomp} 
\usepackage{lmodern} 
\usepackage[T1]{fontenc}

\begin{document}

\title{A hydrogen leak-tight, transparent cryogenic sample container for ultracold-neutron transmission measurements}

\author{Stefan D\"{o}ge}
\email[Corresponding author Stefan Doege: ]{stefan.doege@tum.de}

\affiliation{Institut Laue--Langevin, 71 avenue des Martyrs, F-38042 Grenoble Cedex 9, France}
\affiliation{Physik-Department, Technische Universit\"{a}t M\"{u}nchen, D-85748 Garching, Germany}
\affiliation{\'{E}cole Doctorale de Physique, Universit\'{e} Grenoble Alpes, F-38402 Saint-Martin-d'H\`{e}res, France}

\author{J\"{u}rgen Hingerl}
\affiliation{Institut Laue--Langevin, 71 avenue des Martyrs, F-38042 Grenoble Cedex 9, France}
\affiliation{Physik-Department, Technische Universit\"{a}t M\"{u}nchen, D-85748 Garching, Germany}

%\date{}

\begin{abstract}
The improvement of the number of extractable ultracold neutrons (UCNs) from converters based on solid deuterium (sD$_2$) crystals requires a good understanding of UCN transport and how the crystal's morphology influences its transparency to UCNs. Measurements of the UCN transmission through cryogenic liquids and solids of interest, such as hydrogen (H$_2$) and deuterium (D$_2$), require sample containers with thin, highly polished and optically transparent windows and a well defined sample thickness. One of the most difficult sealing problems is that of light gases like hydrogen and helium at low temperatures against a high vacuum. Here we report on the design of a sample container with two $1\,\text{mm}$ thin amorphous silica windows cold-welded to aluminum clamps using indium wire gaskets, in order to form a simple, reusable and hydrogen-tight cryogenic seal. The container meets the above-mentioned requirements and withstands up to 2~bar hydrogen gas pressure against isolation vacuum in the range of $10^{-5}$ to $10^{-7}$~mbar at temperatures down to 4.5~K. Additionally, photographs of the crystallization process are shown and discussed.
\\
\\
Published as: Review of Scientific Instruments 89, 033903 on 6 March 2018 \\ \url{https://doi.org/10.1063/1.4996296}.
\end{abstract}

%\pacs{07.20.Mc, 61.05.F-, 78.20.-e, 28.20.Cz}

\maketitle 

\section{Introduction}

The scattering of thermal and cold neutrons from cryogenic liquids and solids is a well established experimental technique. Most of the time, sample containers and their neutron beam windows are machined from aluminum alloy, because of its favorable post-irradiation behavior (short half-life of 2.5 min of the $^{28}$Al isotope, which is created by $^{27}$Al capturing a neutron), low neutron scattering and absorption cross section, and easy workability. Thermal and cold neutrons have wavelengths of 1 to 10\,\AA ngstr\"{o}m and are therefore practically insensitive to the aluminum's neutron-optical potential, material inhomogeneities and surface roughness.

Slow neutrons -- especially ultracold neutrons (UCNs), with a velocity of only a few meters per second and a wavelength of several hundred \AA ngstr\"{o}m -- are, however, sensitive to surface roughness\cite{steyerl:1972}, material\cite{steyerl:1975-turbine,pokotilovski:2011} and magnetic inhomogeneities\cite{steyerl:1976}. Besides that, aluminum windows are not optically transparent and thus do not allow for an on-line control of the sample. In UCN applications, the advantage of the low neutron-optical potential of aluminum (54 neV) is more than offset by the drawbacks that the use of this material entails.

In this article, we present an improved sample container design which addresses the issues of previous sample containers and permits reliable UCN transmission measurements on cryogenic crystals, such as solid deuterium (sD$_2$). Measurements of this kind are needed to interpret the performance of operating sD$_2$-based UCN sources worldwide\cite{saunders:2013,karch:2014,lauss:2014}.

\section{The Need for an Improved Sample Container}

If one wants to measure the transmission of ultracold neutrons through a cryogenic liquid or solid, the windows of the sample container need to be as highly polished as possible (center-line average roughness $R_\text{a} < 10\,\text{\AA}$) in order to minimize undesired scattering from surfaces. Machined, rough-surface aluminum windows are fairly easy to make and have been used in UCN transmission experiments before \cite{atchison:2005-liq,atchison:2005-sol,lavelle:2010,doege:2015}. As our tests with single-side polished aluminum windows and unpolished as well as polished aluminum foils ($R_\text{a}$ in the range of a few $\mu \text{m}$) have shown, they are not suitable for UCN transmission experiments because of significant UCN scattering from the vacuum--aluminum and aluminum--sample interfaces due to surface roughness.

In addition, thin aluminum windows in the range of $0.15$ to $0.3\,\text{mm}$ tend to bulge at a pressure difference of about one bar. This results in a poorly defined sample thickness, which translates directly to a large error in the scattering cross section. Atchison et al. \cite{atchison:2005-liq,atchison:2005-sol} used an initial sample thickness of 10.0\,mm. However, after bulging of the windows, computer simulations suggested an ``effective thickness'' of 11.1\,mm. Considering the relation of sample thickness $d$ and the total cross section $\sigma_\text{tot}$ in the transmission equation (eq.~\ref{eq:transmission}, where $I_0$ is the transmitted UCN flux through an empty sample container, and $I(d)$ is the transmitted UCN flux through a sample of thickness $d$, $N_\text{v}$ is the molecular number density of the sample), one immediately understands the importance of a well defined sample thickness.

\begin{equation}
\frac{I(d)}{I_0} = \text{e}^{-N_\text{v} \sigma_\text{tot} d}\label{eq:transmission}
\end{equation}

Besides exhibiting a low surface roughness, sample container windows for UCN transmission experiments should be made from a material with low absorption cross section to maximize the neutron flux to the sample, and with a low neutron-optical potential\cite{ignatovich:1990,golub:1991} to transmit UCNs with an as low as possible energy. All materials are practically impervious to UCNs with a kinetic energy below their respective neutron-optical potential. Materials of choice are thus silicon, transparent vitreous silica and synthetic quartz (SiO$_2$; naming convention suggested by Laufer\cite{laufer:1965} to clarify the naming variations of quartz\cite{agricola:1546}), and sapphire (Al$_2$O$_3$). The latter three are optically transparent and allow for an observation of the condensation and crystal growth processes in the sample container along the neutron beam axis. Amorphous silica has the unique advantage of not producing small-angle scattering inside the material\cite{roth:1977}. In our sample containers we used transparent vitreous silica wafers purchased from Plan~Optik~AG, Elsoff, Germany. All of the following reported results were obtained using sample containers with these wafers.

\begin{figure}[!h]
\includegraphics[width=1.00\columnwidth]{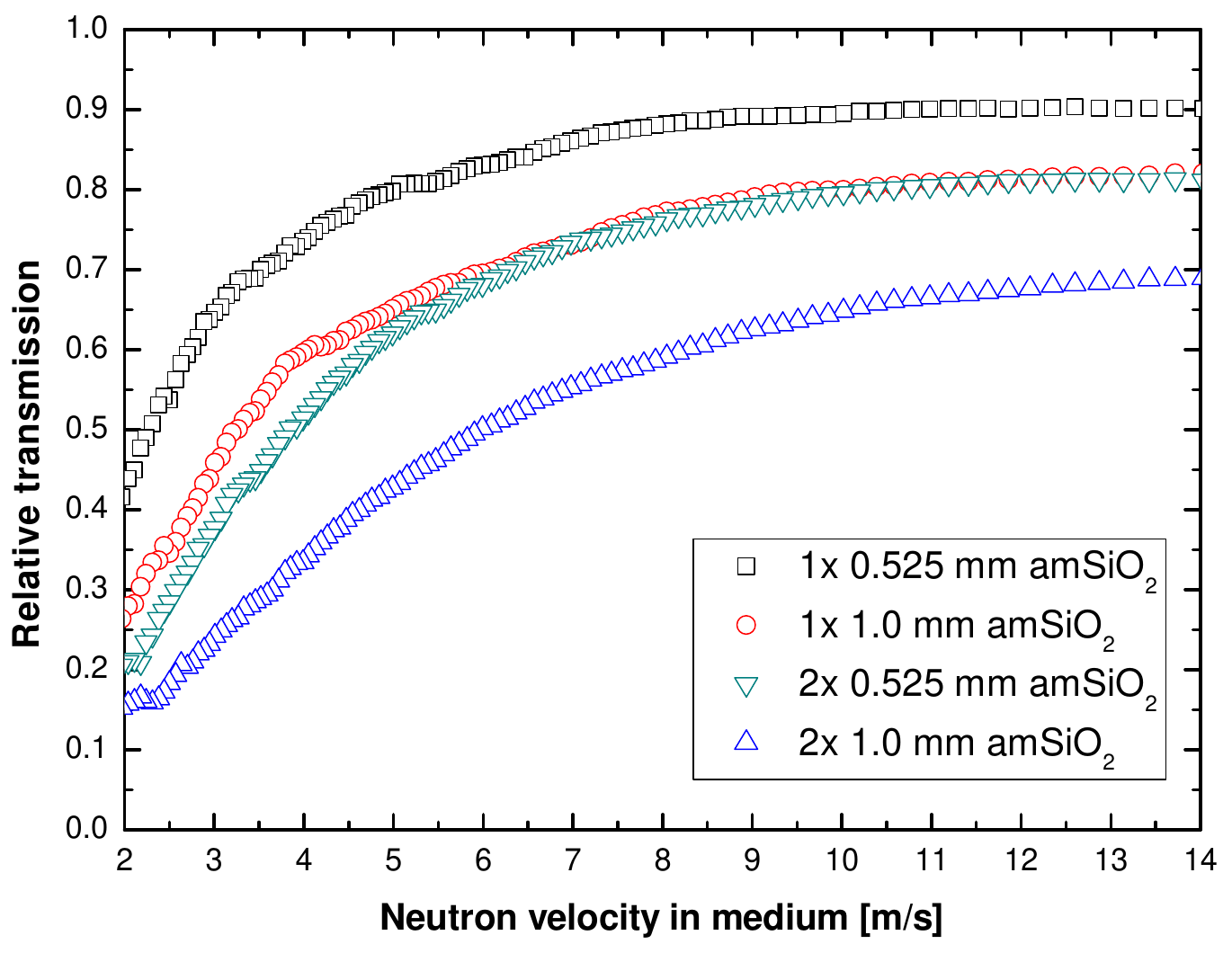}
\caption[Neutronic perormance of silica windows]{\label{fig:neutronics-silica}(Color online). Transmission of UCNs through one and two amorphous silica windows of 1.0 mm and 0.525 mm thickness at room temperature relative to the direct UCN beam (T = I/I$_0$). The blue triangles ({\color{blue}$\triangle$}) represent the final (empty) sample container. This UCN transmission measurement was carried out at the PF2-EDM beamline at Institut Laue--Langevin (ILL).}
\end{figure}

The fact that the surfaces of the silica windows have a negligible influence on the measured UCN transmission is demonstrated by the virtually identical transmissivity through one $d=1.0\,\text{mm}$ window and two $d=0.525\,\text{mm}$ windows (see Fig.~\ref{fig:neutronics-silica}). If the surfaces had a large impact, four vacuum--silica interfaces would transmit substantially less UCNs than two such interfaces. The calculated neutron-optical potential\cite{ignatovich:1990,golub:1991,sears:1992} of our amorphous silica windows based on the volumetric mass density provided by the manufacturer ($\rho = 2.203$ g/cm$^{3}$), is 90.6\,neV at room temperature and the same at cryo-temperatures due to a volume contraction of less than one per mille.

Since the aforementioned suitable materials are commonly supplied as flat wafers and it is very difficult to make them into one-piece structures (like a flat window plus clamp with screw holes) that would fit into our sample container\cite{doege:2015}, we had to develop a hydrogen-tight seal to join the flat wafers and the clamps made from aluminum alloy AlMg3 (AA5754).

\section{Design Requirements of the Sample Container and Optical Access}

A thick and therefore mechanically strong glass slab or collar does not present an obstacle in optical applications. By contrast, in our case the glass windows need to be as thin as possible to minimize the absorption of UCNs. First we tried $0.5\,\text{mm}$ thick vitreous silica windows in our sample container, but they imploded in a vacuum test at about 800\,mbar pressure difference. Same-size windows, but with a thickness of $1.0\,\text{mm}$, withstood a pressure difference of 1\,bar and more, and thus became our window material of choice.

In brief, the design requirements were:
\begin{itemize}
  \item highly polished and thin windows ($d=1\,\text{mm}$), the windows need to be easily removable
	\item optically transparent windows with a low neutron-optical potential and of high purity (no scattering length density inhomogeneities in the material)
	\item vacuum seal needs to be easily demountable and hydrogen-tight down to 4.5 K
	\item sample thickness needs to be well defined, the same across the entire sample area, and adjustable over a wide range within the same sample container
	\item sample container needs to withstand $2\,\text{bar}$ hydrogen pressure against high vacuum with a maximum pressure of $10^{-3}\,\text{mbar}$, preferably in the range of $10^{-5}$ to $10^{-7}\,\text{mbar}$, in order for the closed-cycle refrigerator to work smoothly and to minimize UCN up-scattering on residual gas molecules.
\end{itemize}

\begin{figure}
\includegraphics[width=1.00\columnwidth]{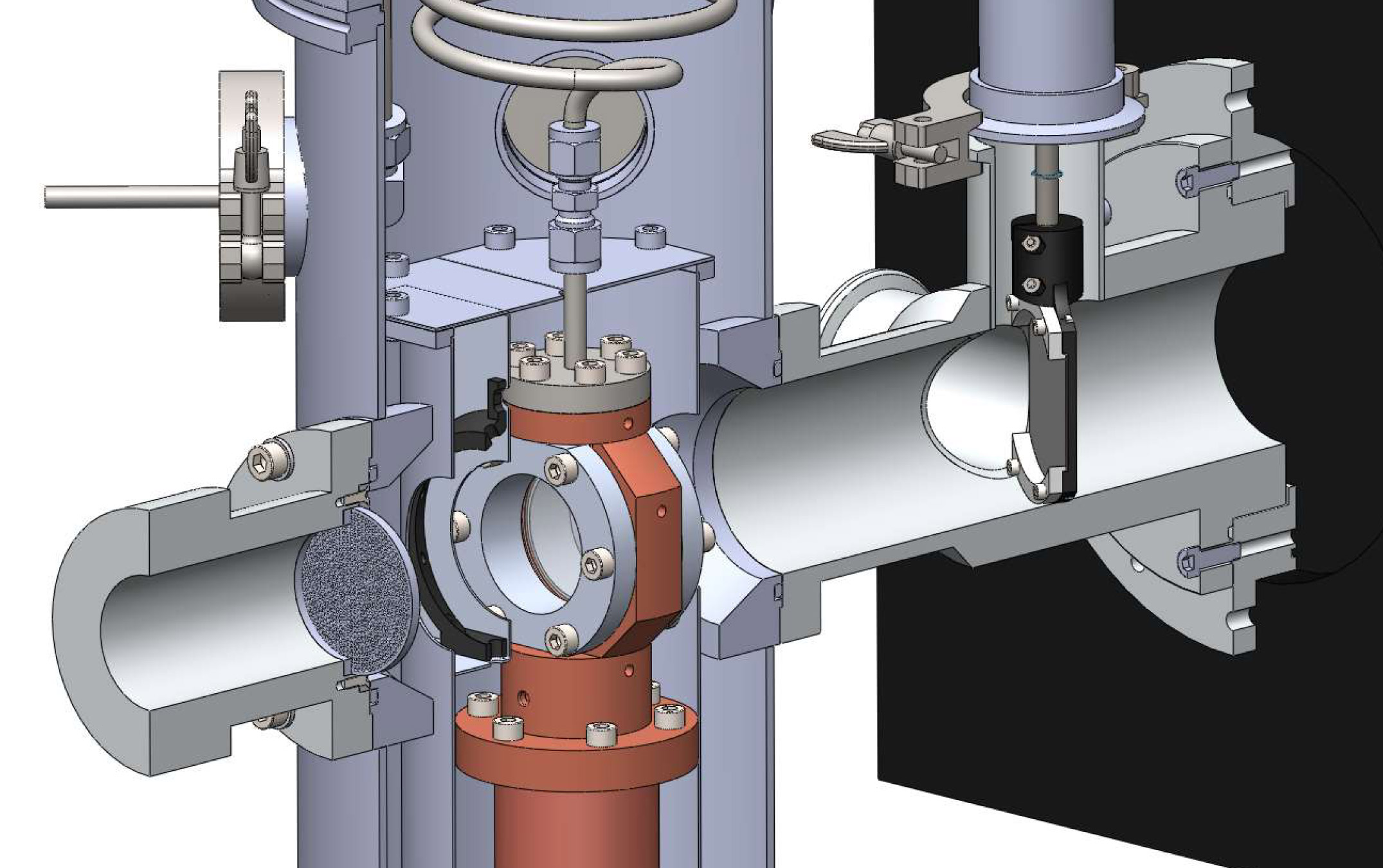}% Here is how to import EPS art
\caption[Cryostat]{\label{cryostat-oblique}(Color online). The cryostat is shown with the transparent sample container mounted. From left to right the components are: neutron guide adapter, neutron collimator, outer cryostat body, cryo-shield connected to the 1$^\text{st}$ cooling stage of the cold-head, LED support and heat reflector (not shown), transparent sample container (with gas feed line) mounted on the 2$^\text{nd}$ cooling stage, neutron guide with horizontal viewport (glass flange) and vertical vacuum feed-through for the shaft and mirror (only black plastic mirror support is shown), and the neutron detector (black box on the right).}
\end{figure}

The foot of the sample container was designed to be mounted onto the two-stage cold-head of the cryostat described in D{\"o}ge et al.\cite{doege:2015} with a number of modifications: The stainless steel neutron guides facing directly the sample container were removed. An illumination unit with nine white LEDs was mounted on one side of the sample container (see Fig.~\ref{cryostat-oblique}). The LED support had a recess for a 50.8 mm double-side polished undoped silicon wafer that served as a reflector for heat radiation coming from the neutron guide. The LED support, and with it the heat reflector, were thermally connected to the cryo-shield on the 1$^\text{st}$ stage of the cold head (40 K). The uncovered side of the container could be observed optically through a glass window in the neutron guide and a polished-silicon mirror that could be raised and lowered into the neutron beam through a vacuum feed-through. The mirror was attached to a shaft that could be turned around 360$^\circ$, which allowed us to find the best mirror position for sample observation.

\section{Indium Seal}

Since the sample container and its windows needed to be sealed against gaseous and liquid hydrogen at temperatures as low as 5\,K, the use of rubber O-ring gaskets was not possible. At low temperatures they usually become brittle and leak. Only a few elaborate designs can work with rubber gaskets at low temperatures\cite{robbins:1964}. The seal for our purposes needed not only to withstand cryogenic temperatures, but it also had to wet the sealing surfaces nearly perfectly. As the smallest molecule in existence, hydrogen is very mobile and can hence escape through even the tiniest scratches in the sealing surfaces. Therefore, the gasket material selected was an indium metal wire gasket (in the literature sometimes referred to as indium O-ring).

Presumably the first mention\cite{rose:1944} of indium adhering to glass due to its high wettability was made in 1944. Several years later, in experiments to solder indium metal onto thin films on glass substrates, the adhesiveness of indium to a variety of materials was determined by Belser\cite{belser:1954:p180}. Among them were several materials which are of importance to slow neutron scattering -- as windows for sample holders: silicon, quartz, aluminum oxide; structural materials: aluminum and copper; and as neutron absorbers: cadmium, titanium and lead. Belser soldered at temperatures around 160\textdegree C to make the adhesive connection between indium and these materials. Our sample container's geometry, however, did not allow for a hot treatment.

Cold-welding an indium seal, i.e. applying only pressure at room temperature to fuse the ends of an indium wire to one another and the whole one-piece ring to a glass surface, appears to have first been reported by Edwards\cite{edwards:1956}.
Following this pioneering work, many designs of cold-welded indium wire seals have been published; mostly for metal-to-metal joints, but also for metal-to-glass joints. The term ``glass'' here is meant in a broader sense to refer to optically transparent materials and includes quartz, vitreous silica, sapphire, and similar materials. The main advances in the field of metal-to-glass joints that are hydrogen and helium-tight, have been made in the 1950s-1960s\cite{edwards:1956,willis:1958,horwitz:1961,smith:1962,lipsett:1966}, among others to seal a large liquid hydrogen bubble chamber\cite{lucas:1959}, and in the 1980s\cite{abraham:1975,turkington:1984,lim:1986,haycock:1990}. A substantial overview of published indium seal designs was given by Turkington and Harris-Lowe\cite{turkington:1984}. All previous designs that describe metal-to-glass joints use, however, flat glass slabs of appreciable thickness -- in the range of several millimeters to centimeters -- and of corresponding mechanical strength.

\section{Final Design}

Two particular restrictions to the design of our sample container made it difficult to achieve good vacuum tightness: (i) The need to have thin windows in combination with a large diameter of 43.8\,mm prohibited the exertion of excessive clamping force on the silica windows. (ii) Depositing a wetting agent like indium, nichrome or platinum onto the rims of the glass flats to improve contact with the indium gasket, as some experimenters have done\cite{lipsett:1966,abraham:1975,lim:1986}, was not desirable in our case because of frequent demounting of the sample container and some glass breakage that occurred. This meant that the bare indium gasket alone had to provide the required vacuum tightness.

We first experimented with unchamfered aluminum compression clamps for the indium seal and were able to build containers tight enough to condense air and hydrogen into them. However, these seals were not stable over time and proved to fail at overpressures of a few hundred millibars.

The solution to this problem was to machine a $1\,\text{mm}$ deep 45\textdegree\, chamfer on the outside of the compression clamp's rim that protrudes into the sample container. The chamfer created a void for the indium to creep into, which could then evenly distribute the clamping force over the whole edge of the silica window. The chamfer greatly improved reproducibility and reliability of the seal.

If the seal were classified according to Lim\cite{lim:1986}, it would be called a partially trapped O-ring seal.

In UCN transmission experiments it is important to measure samples of two or more well defined different thicknesses so as to separate bulk scattering from scattering at the surface or the sample--window interface. Our copper sample container employs aluminum spacer rings of variable thickness which, combined with a series of different pressure clamps adapted to them, enable quick adjustment of the sample thickness. The two aluminum pressure clamps require a total of four indium wire seals -- two inner seals (clamp to silica window) and two outer seals (container body to clamp mating surface).

The clamping force is provided by six equally spaced stainless-steel hexagon socket head cap screws (size M5, 16 mm thread) on each of the two aluminum clamps. To prevent the screws from loosening at cryogenic temperatures, split lock washers are placed between them and the pressure clamps. As the clamps press evenly against the inner and outer indium wire gaskets, the silica windows are held at a constant distance by 2\,mm wide (i.e. outside diameter minus inside diameter divided by 2) spacer rings made from aluminum that support the edge of the windows. That distance remains constant even under high pressure. At the top of the spacer ring there is a gap for the gas inlet. The edges of the aluminum spacer rings need to be smoothly rounded off, in order to avoid extreme local strain on the silica windows which can lead to their cracking.

\begin{figure}
\includegraphics[width=1.00\columnwidth]{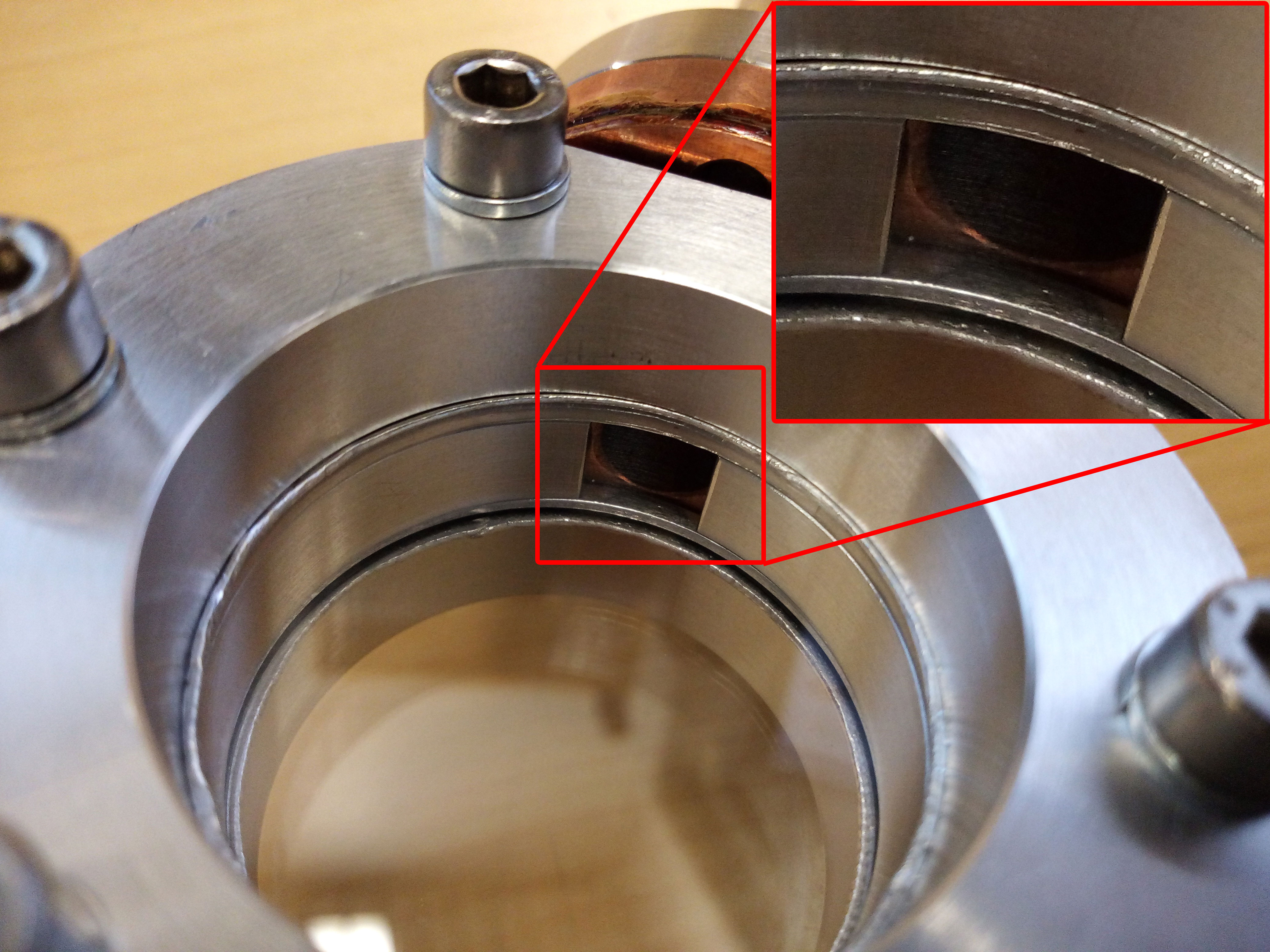}
\caption[Sample container detail]{\label{sample_container_detail}(Color online). Close-up view of the assembled sample container showing the gas inlet and the gap in the central aluminum spacer ring. The layers from the top as shown in the inset are: aluminum clamp, indium wire gasket, silica window, first spacer ring ($d=1.5$~mm), second spacer ring ($d=9$~mm) with 9mm gas inlet gap, third spacer ring ($d=1$~mm), silica window, indium wire gasket, aluminum clamp.}
\end{figure}

Although spacer rings with a gas inlet gap as large as the diameter of the gas inlet itself would be favorable (9~mm in our case), we found the optimal gap width to be 5 mm. Rings with a 9 mm gap made the silica windows crack due to the larger unsupported fringe area of the windows. It is, however, possible to use rings with 5 mm gaps or no gaps as direct supports for the silica windows and a center ring with a wider gap to allow for a larger gas flow area (see Fig.~\ref{sample_container_detail}).

\begin{figure*}
\includegraphics[width=2.00\columnwidth]{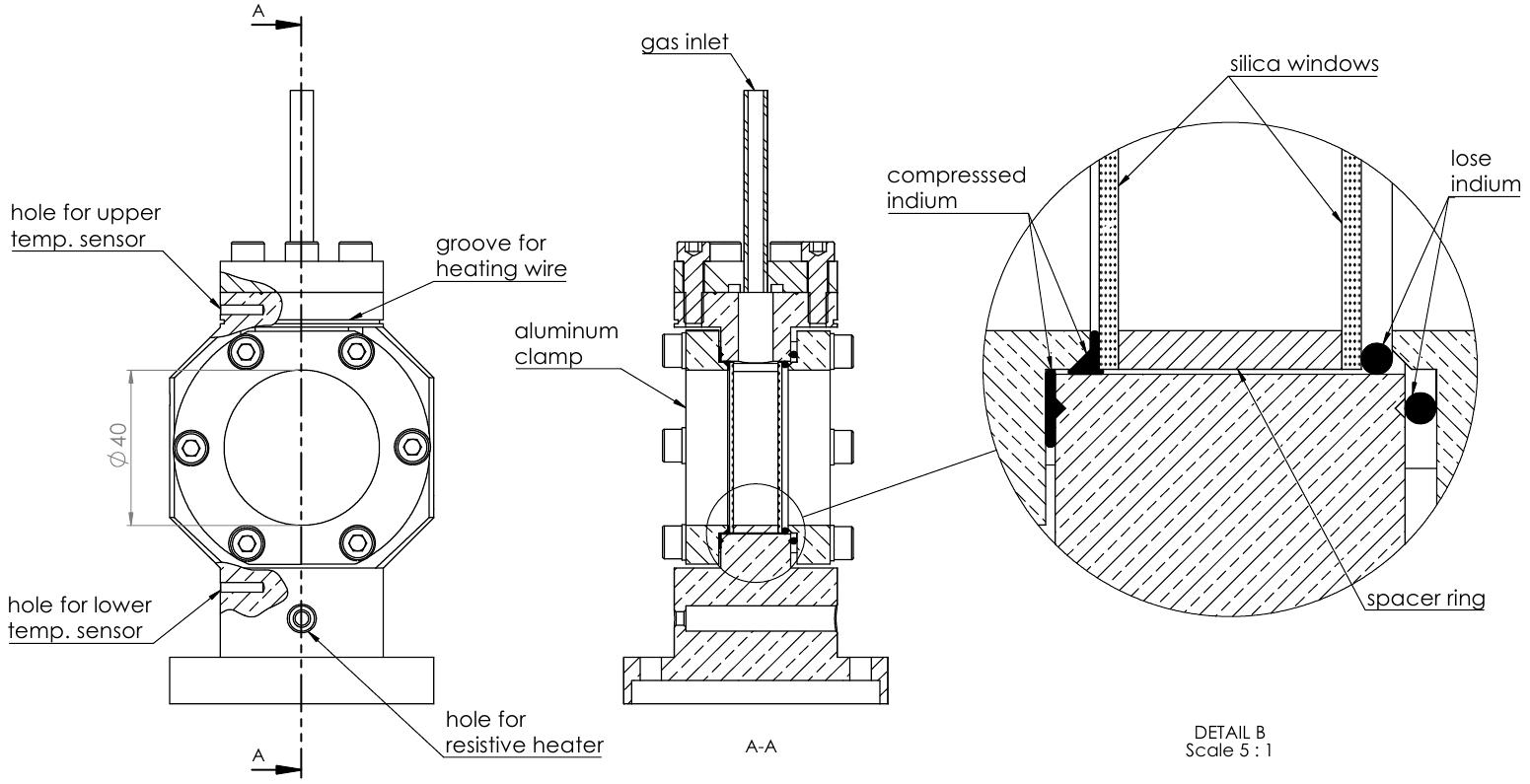}
\caption[Sample container]{\label{sample_container}Front, side and detailed view of the sample container; each of the two aluminum compression clamps is affixed to the copper sample holder by six stainless steel hex screws (size M5, 16 mm thread). Split lock washers or copper--beryllium washers are placed under the screw heads to prevent loosening of the assembly at cryogenic temperatures. The copper container has one $1\,\text{mm}$ deep angular groove at 45\textdegree\, for the outer shallow-groove indium seal\cite{fraser:1962} on each mating surface. The outside of the rim of the aluminum clamp, which protrudes into the copper container, has a $1\,\text{mm}\times 45$\textdegree\, chamfer in order to tightly press the indium against the silica window and the copper container body. The two windows are kept at a precisely determined distance from each other by an aluminum spacer ring between them, which has a thickness of $11.5\,\text{mm}$ in this figure. In the sectional view A--A and the detailed view B, the left-side indium wire gaskets are compressed, the right side shows uncompressed indium wire of $1.5\,\text{mm}$ diameter. This drawing is to scale, the dimension (40 mm) of the aluminum clamps's inner diameter is indicated on the left.}
\end{figure*}

The length of the compression clamp protruding into the sample container and the thickness of the spacer rings were designed such that the inner indium wire seals (clamp to silica window) were compressed to a thickness of $0.50$ to $0.60$\,mm, and the outer seals (container body to clamp mating surface) to $0.45$ to $0.50$\,mm. With the initial indium wire thickness of 1.5\,mm this meant a final gasket compression to 1/3 of the original thickness (see Fig.~\ref{sample_container}, detail B).

As was pointed out by Turkington and Harris-Lowe\cite{turkington:1984}, to ensure high seal tightness, the surface that is cold-welded to the indium gasket, needs to be either untreated, i.e. utilized as it comes off the lathe, or polished to less than 5\,$\mu$m roughness. In our seal the vitreous silica wafers were highly polished ($R_\text{a} < 3\,\AA$, as determined by atomic force microscopy) and the bonding of indium wire to the wafer could be seen after disassembly of the sample container. The aluminum compression clamps were used with the surface finish as they came off the lathe, i.e. with microscopic spiral lines causing a surface roughness of $R_\text{a} \approx 7\,\mu \text{m}$. The inside surface of the copper container body was equally untreated.

For sample heating, the container is equipped with two heaters, one 0.1\,mm diameter constantan resistive wire recessed in a groove around the top, and one heating resistor in the base of the container. Two Cernox thin film resistance cryogenic temperature sensors take the temperature readings just above and just below the sample volume. They are inserted into holes drilled into the container body (see Fig.~\ref{sample_container}, detail A).

\section{Assembly and leak testing at room temperature}

Before inserting the aluminum spacer rings, the silica windows, the indium wire gaskets and the aluminum clamps into the sample container, all these items were cleaned. The aluminum parts and the sample container itself were wiped with a lint-free wipe soaked in pure ethanol. The silica windows were pre-cleaned by the manufacturer in a process comprising rinses with a sodium hydroxide solution followed by a phosphoric acid solution. The acid was removed using de-ionized water and the windows were then blow-dried with air. We wiped each of them with ethanol and then blow-dried them with oil-free air to remove any lint from the surface.

Making indium wire more adhesive is achieved by removing the \textasciitilde 100\, \AA{ }oxide layer that forms naturally on the surface. A method proposed by the Indium Corporation\cite{indium-corp} (degreasing, wiping with 10\,wt.-\% hydrochloric acid, rinsing with de-ionized water and acetone) improved the adhesiveness of the indium wire, but proved to be rather time-consuming. Instead, we removed the oxide layer mechanically by scraping it off with a knife that was pulled gently over the wire against the direction of the cutting edge. This treatment improved the adhesiveness to the same extent. The indium wire used in this joint had been recycled from scrap indium of $>99.999$\,mol-\% original purity by the Cryogenics Service of the ILL. An inductively coupled plasma mass spectrometry (ICP-MS) analysis of the wire showed a purity of $99.7$\,mol-\%, with the main contaminants being Pb, Sn, Ag, Na, and Cd. This is the standard gasket material for cryostats at ILL. Neither the indium wire gaskets, nor the mating surfaces were coated with solder flux or grease, as was the case for some previous seal designs\cite{lipsett:1966}.

Indium wire rings of 43.5\,mm inside diameter for the two inner seals and of 47.0\,mm for the two outer seals were preformed from 1.5\,mm diameter indium wire. The wire's butting ends were beveled by cutting them diagonally with a sharp blade at an angle of about 45\textdegree\, to 60\textdegree . The exact angle did not seem to have any influence on the seal tightness as long as the cut faces were properly aligned and pressed against each other.

Assembly of the sample container commenced by attaching one of the aluminum compression clamps and the corresponding outer indium wire gasket to the sample container body (inner diameter of 44.5 mm) with 6 hex screws. Then the container was turned around such that all other inserted parts could rest on the protruding rim of that compression clamp. Next the first inner indium wire gasket was inserted into the ledge formed by the rim of the clamp and the inner wall of the sample container. After the indium wire gasket had been put in place and leveled, a vitreous silica window and one or more spacer rings made from aluminum were inserted. At this stage it had to be ensured that the gap in the spacer ring was properly aligned with the gas inlet port in the top of the container. The following items were then inserted: the second vitreous silica window and the second inner and outer indium wire gaskets. The last assembly step was to carefully slip the second compression clamp into the orifice of the sample container and to press lightly but equally against both the inner and outer indium wire gaskets.

After all parts were in place, the container could be returned to an upright position. The stainless steel screws on the pressure clamps were then tightened step by step in a criss-cross pattern to ensure a uniform pressure distribution on the gaskets and the silica windows. The seal was formed as all four indium wire gaskets were gradually compressed. %see Lim, ref. 11

Since the indium continued to flow slowly each time the screws were tightened, we allowed about five minutes before the next round of tightening.

In the thin-sample configuration ($d < 4.5$ mm), the gas inlet ($d=9$ mm) overlaps the sandwich of indium wire, window, spacer, window, indium wire and the sample volume is connected to the annular space of about 0.09\,cm$^3$ between the two indium seals through a small gap between the wall of the gas inlet and the compressed indium wire (see the sectional view A--A in Fig.~\ref{sample_container}). During evacuation of the sample container, the annular space is evacuated as well. However, in the thick-sample configuration ($d \geq 4.5$ mm) the annular space is completely trapped between the inner and outer indium seals and can therefore not be evacuated. Air that gets trapped there during assembly of the container, cannot escape. Since the sample container is well evacuated and cooled down to cryogenic temperatures before inserting hydrogen and other gases into the sample volume, this air solidifies inside the annular space and can therefore not contaminate the sample. The trapped air in the annular space did in no way compromise the functionality of the sample container described here. In future designs, a modified shape of the gas inlet can connect the annular space to the sample volume, which can thus be vented during evacuation of the sample container.

The leak rate out of the sample container into the isolation vacuum, where some of the surfaces are at low temperatures, had to be as low as possible. Only below $10^{-1}\,\text{mbar}$ are the up-scattering losses of ultracold neutrons negligible\cite{doege:2019-phd} and only well below $10^{-2}\,\text{mbar}$ does a cold-head (closed-cycle helium refrigerator) work properly. The isolation vacuum was evacuated by a pre-pump and a turbomolecular pump used in series. A desirable permanent isolation vacuum with the sample container installed and filled is below $10^{-5}\,\text{mbar}$.

During extended operation of the cryostat at cryogenic temperatures, gas from small leaks into the isolation vacuum can potentially freeze out on the cold-head and suddenly evaporate during warm-up of the cryostat. To prevent the risk of pressure peaks on the experimental equipment and the UCN source, the cryostat was equipped with an overpressure break foil. In the unlikely event of a sudden pressure increase, it would have opened up and released the gas into the atmosphere. The installation of the break foil was in compliance with the safety regulations in force at Institut Laue--Langevin.

Since the installation of the sample container into the cryostat and the subsequent vacuum pumping and cool-down are quite time-consuming, we developed a leak testing method, based on a pressure build-up measurement at room temperature, from which we could reliably deduce the sample container's performance at cold temperatures. After complete assembly and the tightening of the indium seals, we leak-tested the containers by evacuating them down to the $10^{-2}\,\text{mbar}$ range using an oil-free piston pre-vacuum pump against 1\,bar atmospheric pressure. The screws on the pressure clamps were then continuously tightened until the rate of air leaking into the sample container was below $Q_\text{L}^\text{thick}=3.4 \times 10^{-5}\,\text{mbar}\times \text{L}/\text{s}$ and $Q_\text{L}^\text{thin}=9.5 \times 10^{-5}\,\text{mbar}\times \text{L}/\text{s}$ for thicker ($d\geq 4.5\,\text{mm}$) and thinner sample thicknesses, respectively. These values have been corrected for the leak rate of the gas test stand and were calculated over a time period of ten minutes, which is about the time it takes to fill the container with a cryogenic liquid.

After the vacuum leak test, the sample container was vented and the indium wire was left to creep for 12 to 24 hours. Then the vacuum leak test was repeated and the screws gradually tightened until the leak was again at or below $Q_\text{L}$. In this last step, a torque screwdriver may be useful. The maximum final torque applied to the screws was between 0.8 to 1.0 Nm. The last leak test and tightening should be done only shortly before mounting the sample container into the cryostat.

To obtain a general idea of the seal tightness against hydrogen, overpressure tests were performed with a similarly light and viscous, but safer gas -- helium. The sample container and the gas system were filled with 1600\, mbar of helium against 1 bar of atmospheric pressure. These tests usually lasted for one or two hours. The leak rate of helium out of the sample container and into the atmosphere was typically $\leq 2 \times 10^{-4}\,\text{mbar}\times \text{L}/\text{s}$ (corrected for gas test stand leakage) for containers that had been successfully air-leak tested in a pressure build-up measurement.

Sample containers assembled and tested in this manner remained deuterium-tight for four temperature cycles between 5 K and room temperature with deuterium condensed into and evaporated out of the container each time. The tests were discontinued at that point with the container fully functioning, because frequent cycling between room and cryogenic temperatures is not required in our application. As it stands, the container would have likely remained intact during more cyclings.

\section{Observations at Cryogenic Temperatures}

The first cryogenic tests using this sample container were done by freezing a deuterium crystal out of the gas phase. In order to have the warm gas freeze at all, the container had to be as cold as possible as opposed to being kept just below the freezing point. As the warm gas flowed into the container, it started freezing out on the bottom, but to a large degree also on the side walls and inside the gas inlet -- copper parts that were well thermalized to the cold-head temperature. In all attempts the gas inlet froze over before the crystal in the sample container had grown to a sufficient size. That is why freezing from the gas phase was not pursued any further and the crystal was instead grown from the liquid. 

Other experimenters, for example Lavelle et al.\cite{lavelle:2010}, encountered a similar problem. Due to the use of an optically non-transparent sample container, they were not able to verify whether solid deuterium (sD$_2$) in a half-filled container rested only in the lower half or whether part of it froze out on the side and upper walls of the container. Our observations support the latter conjecture.

When cryogenic liquids (H$_2$, D$_2$, neon) close to the triple point were kept in our sample container, the container and its indium gaskets had to withstand pressure differentials of up to 2\,bar. A hydrogen pressure of about 1.5\,bar in the sample container increased the pressure of the insulating vacuum of the cryostat to the 10$^{-5}$\,mbar range. The seals withstood higher pressures for a few minutes. The pressure inside the container was measured on the gas fill line just outside of the cryostat\cite{doege:2015}.

\begin{figure*}%[!h]
\includegraphics[width=1.75\columnwidth]{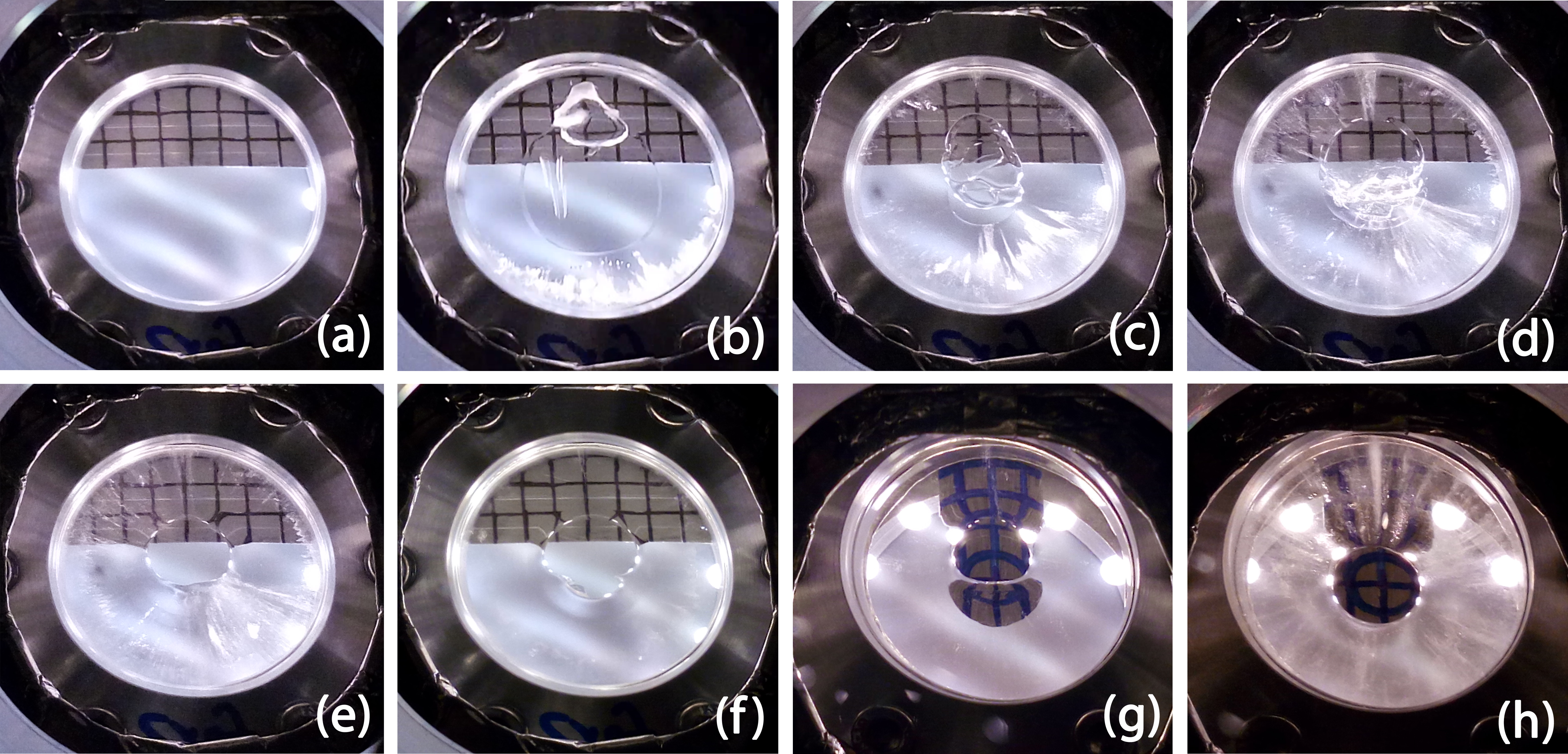}
\caption[Hydrogen sample]{\label{fig:hydrogen-sample}(Color online). View of the sample container along the neutron beam axis as seen through viewport and mirror shown in Fig.~\ref{cryostat-oblique}. Parts of the sample container are blinded out for ultracold neutrons by $0.5\,\text{mm}$ thick cadmium absorbers. The marking lines on the absorbers are for reference and are $5\,\text{mm}$ apart from each other. The sample in (b-f) is $4.5\,\text{mm}$ \textit{para}-hydrogen, in (g-h) $6.5\,\text{mm}$ \textit{ortho}-deuterium. The images above show (a) an empty sample container with a straight-edge absorber; (b) a liquid-filled \textit{para}-hydrogen crystal ring with bubble formation in the liquid phase; (c) freshly frozen solid \textit{para}-hydrogen, irregular solid--liquid--vacuum phase boundaries (``snow'') in the center of the sample container, visible radial streaks; (d) the previous crystal after one thaw--refreeze cycle; (e) the previous crystal after about 10 thaw--refreeze cycles, where the irregular phase boundaries have disappeared and a void with a smooth surface has formed; (f) the previous crystal after 15 hours at constant temperature ($T=9\,\text{K}$) and one short thaw--refreeze cycle, the radial marks have disappeared as they were located not in the crystal bulk, but only on the crystal--window interface; (g) a frozen and cycled \textit{ortho}-deuterium crystal at $T=14.5\,\text{K}$ with a flap-shaped absorber; (h) the previous crystal after cooling to $T=10\,\text{K}$.}
\end{figure*}

After the first crystal growing tests in the transparent sample container it quickly became apparent that substances with a higher density in the solid than in the liquid phase will always form bubbles in the freezing process (see Fig.~\ref{fig:hydrogen-sample}). In the case of deuterium the density increases by 12\,\% during the phase transition from liquid to solid, for hydrogen the increase is 11\,\%\cite{souers:1986}. The reason for bubble formation is that the crystal solidifies out of the liquid radially, starting at the inner wall of the sample container and growing inward. The fastest growth of the crystal takes place at the bottom of the sample container, where most heat is removed from the sample by the cold-head. But even on the sides and the top of the sample container, where the temperature is about 1\,K higher than at the bottom due to the inflow of warm gas, the liquid starts freezing. When about one third to one half of the sample is frozen, the gas inlet at the top of the sample container freezes over and no additional liquid can enter the sample container. This cannot be avoided, even by heating the top of the sample container with 4 to 5 watts, because the power required to keep the gas inlet open would heat up the entire sample to above the freezing point.

After the gas inlet has frozen over and a liquid-filled crystal ring around the inner wall of the sample container has formed, there is a point at which gas bubbles form inside the liquid as the freezing process continues. They rise to the top of the liquid reservoir and are pushed downward by the advancing crystal ring. When all liquid has frozen, the center of the sample container is filled with an irregularly shaped solid--vacuum phase boundary. This is shown in Fig.~\ref{fig:hydrogen-sample} (c) -- albeit with a minimal amount of suspended liquid. One might call this phase ``snow''. UCNs are highly sensitive to phase boundaries and rough surfaces where the neutron-optical potentials of both phases differ by more than \textasciitilde 10\,neV. As the highest neutron flux is very much centered in UCN guide tubes, this snow is right inside the UCN beam and significantly distorts the scattering pattern.

The only way we found to circumvent these complications was to minimize the snow-covered area by employing thaw--refreeze cycles. After about ten such cycles (see Fig.~\ref{fig:hydrogen-sample} (e)), during which the snow and the surrounding solid were carefully melted at 0.1\,K above the melting point and then slowly refrozen, we obtained a single void with a smooth surface. The void stretched from one sample container window to the other. Blinding out the void by covering it with a flap-shaped cadmium absorber ($d = 0.5\,\text{mm}$, transmissivity for UCN $\leq 10^{-2}$) and lifting up the sample container by $9\,\text{mm}$ to maximize the UCN flux through the pure crystal below the absorber flap was the only way to obtain UCN transmission data through clear cryogenic crystals (see Fig.~\ref{fig:hydrogen-sample} (g-h)).

This snow-forming behavior was observed in all substances that we froze (hydrogen, deuterium, neon) and at various sample thicknesses (1\,mm, 2\,mm, 4.5\,mm, 6.5\,mm, 11.5\,mm). Although care must be taken when generalizing the behavior of a substance in one specific sample container, it is safe to say that it is very difficult to completely fill any sample container of a few milliliters in volume, such as ours, with a bubble-free crystal.

Fig.~\ref{fig:hydrogen-sample} (e) shows a few radial streaks covering the crystal area. By keeping the crystal just below the melting point and letting heat radiation impinge on the sample container, as well as through melting and refreezing the samples we could establish that these streaks were only present on the crystal--window interface, but not in the bulk of the crystal. A streak-free and absolutely clear \textit{para}-hydrogen crystal is shown in Fig.~\ref{fig:hydrogen-sample} (f).

Whether or not the cryogenic crystal remains in close contact with the vitreous silica window after it has been grown cannot be stated with certainty. However, if it becomes detached, the smooth surface on which it grew (the roughness of silica windows was less than 3\,\AA) and its optical transparency mean that a very smooth crystal surface is highly likely.

The determination of precise UCN total cross sections of cryogenic liquids and solids, especially those of solid deuterium, are of great importance to the planning and improvement of UCN sources based on solid \textit{ortho}-deuterium. The larger the mean free path $\lambda_\text{mfp}=(N_\text{v}\sigma_\text{tot})^{-1}$ of UCNs inside the converter material, the higher the UCN flux extractable from the converter. The mean free path of UCNs, and with it the maximum UCN density inside the converter, depends to a large extent on the sD$_2$ crystal preparation method and holding conditions.

In a previous experiment at the Paul-Scherrer-Institut (PSI)\cite{atchison:2005-sol,brys:2007}, the deuterium crystal was observed perpendicular to the neutron beam axis with blue light from an argon laser. Multiple temperature cyclings between 5\,K and 18\,K significantly deteriorated the transmission of light and UCNs through that crystal. Applying our above observations, it is conceivable that the crystal surface started melting close to 18\,K (even though the temperature sensors were still below the triple point) and then refroze in an uncontrolled fashion. This would have resulted in the gradual formation of a rough crystal surface on the crystal--sapphire interface, causing decreased light transmission, as well as at the crystal--aluminum interface, causing decreased UCN transmission. This scenario is at least a potentially plausible explanation for a significant part of the ``additional isotropic elastic scattering'' that was entirely attributed to crystal imperfections and subtracted from the total scattering cross section before the final publication of results\cite{atchison:2005-sol}.

It should be noted that the silica windows of an empty sample container have a temperature of about 60 K, when the container's body is at 5 K. This is due to the relatively poor heat conductivity of amorphous silica. The corresponding up-scattering cross section for UCNs in amorphous SiO$_2$, calculated using the incoherent approximation and a Debye temperature of 361 K\cite{fukuhara:1997}, is 0.8 barn per SiO$_2$ unit for neutrons with $v = 10$ m/s and can therefore be neglected. As soon as a sample was introduced to the container, the windows adopted the temperature of the sample to within a few 0.1 K.

\section{Conclusions}

We have designed and constructed a hydrogen-tight, easy-to-disassemble sample container for cryo-liquids and cryo-solids with optically transparent windows. It is particularly suited for transmission experiments with ultracold neutrons as it features highly polished window surfaces. The behavior of liquid and solid hydrogen, deuterium and neon inside this sample container has been observed optically and with ultracold neutrons. This new sample container proved indispensable for reliable sample preparation as well as in obtaining scattering cross sections without parasitic effects like scattering on snow and rough surfaces.

Drawing on the insights from this paper, we offered a possible explanation for the ``additional isotropic elastic scattering'' of ultracold neutrons from solid deuterium that was observed in the Ph.D. thesis of T. Bry\'s\cite{brys:2007}.

This paper and a forthcoming publication on ultracold-neutron transmission experiments through liquid and solid hydrogen and deuterium, as well as solid neon, that were performed using the sample container described here, are part of the Ph.D. thesis of Stefan D\"oge\cite{doege:2019-phd}.

\begin{acknowledgments}
We wish to acknowledge the support provided by the mechanical workshops of the Physik-Department of TU M\"{u}nchen and of Institut Laue--Langevin (ILL) in manufacturing the sample container and clamps. For valuable advice we would like to thank Olivier Losserand and Eric Bourgeat-Lami from the ILL Cryogenics Service, Bernhard Lauss from PSI (Villigen/Switzerland) and Christoph Morkel from TU M\"{u}nchen. We also thank Peter Hartung of LMU Munich for the original design of the sample container as described earlier\cite{doege:2015} and Alexander V. Strelkov of JINR (Dubna/Russia) for the kind permission to use his gas test stand at ILL Grenoble. Michael Schneider of Swiss Neutronics is acknowledged for AFM measurements of our silica and aluminum windows. This work received financial support from FRM~II/ Heinz Maier-Leibnitz Zentrum (MLZ), Munich, and from Dr.-Ing. Leonhard-Lorenz-Stiftung, Munich, under grant no. 930/16.
\end{acknowledgments}

\bibliography{Bibliography-arxiv2}

\end{document}